\documentclass[final,5p,times,twocolumn,sort&compress]{elsarticle}

\usepackage{graphicx}

\usepackage{amssymb}

\usepackage[T1]{fontenc}
\usepackage{amsmath}
\usepackage{amssymb}
\usepackage{subcaption}

\biboptions{square}

\newcommand{\abs}[1]{\left|{#1}\right|}
\newcommand{\bkt}[1]{\left({#1}\right)}
\newcommand{\od}[1]{\!\bkt{#1}}

\newcommand{\mean}[1]{\left<{#1}\right>}
\newcommand{\real}[1]{\mathrm{Re}\left\{{#1}\right\}}
\newcommand{\sbkt}[1]{\left[{#1}\right]}
\newcommand{\prob}[1]{\mathcal{P}\!\bkt{#1}}
\newcommand{\tprob}[3][]{\mathcal{P}_{#1}\!\bkt{#2,#3}}
\newcommand{\tprobq}[3][]{\mathcal{P}_{#1}^{\,\mathrm{q}}\!\bkt{#2,#3}}
\newcommand{\probcon}[2]{\mathcal{P}\!\bkt{\left.{#1}\right| #2}}
\newcommand{\intsin}{\mathrm{Si}}

\newcommand{\I}{\mathrm{j}}
\newcommand{\TS}{T_\mathrm{S}}
\newcommand{\DxQ}{\Delta^\mathrm{q}}
\newcommand{\Dy}[1][]{\Delta_\mathrm{#1}}
\newcommand{\xD}[1][]{x^{\mathrm{q}}_{#1}}

\newcommand{\SNR}{\mathrm{SNR}}
\newcommand{\SNRq}{\SNR_\mathrm{q}}
\newcommand{\SNRNL}[1][]{\SNR_\mathrm{nl{#1}}}
\newcommand{\NS}{N_\mathrm{S}}
\newcommand{\Neff}{N_\mathrm{eff}}
\newcommand{\lid}[3]{\int_{#2}^{#3}\mathrm{d} {#1}\,}
\newcommand{\erf}[1]{\mathrm{erf}\od{#1}}
\newcommand{\e}{\mathrm{e}}
\newcommand{\sigmanls}{\sigma_\mathrm{nls}}
\newcommand{\sigmaq}{\sigma_\mathrm{q}}

\newcommand{\sigmaw}[1][]{\sigma_{w{#1}}}
\newcommand{\Zn}{\mathbb{Z}_n}

\newcommand{\secref}[1]{Sec.~\ref{#1}}
\newcommand{\figref}[1]{Fig.~\ref{#1}}

\journal{Signal Processing}

\begin{document}

\begin{frontmatter}



\title{Effective number of samples and pseudo-random nonlinear distortions in digital OFDM coded signal}


\author{Adam Rudzi{\'n}ski}
\ead{adam.rudzinski@ire.pw.edu.pl}
\address{Institute of Radioelectronics, Warsaw University of Technology, ul. Nowowiejska 15/19, 00-665 Warsaw, Poland}

\begin{abstract}
This paper concerns theoretical modeling of degradation of signal with OFDM coding caused by pseudo-random nonlinear
distortions introduced by an analog-to-digital or digital-to-analog converter.
A new quantity, effective number of samples, is defined and used for derivation of accurate expressions for autocorrelation
function and the total power of the distortions. The derivation is based on probabilistic model of the signal
and its transition probability. It is shown, that for digital (discrete and quantized) signals the effective number of samples
replaces the total number of samples and is the proper quantity defining their properties.
\end{abstract}


\end{frontmatter}


\section{Introduction}
In the recent years in an enormous number of telecommunication systems there has been implemented OFDM coding.
This method has been developed already more than 40 years ago \cite{Chang66, Saltzberg67, Weinstein71}, but for several
years has not been used extensively, probably partly due to technical limitations. Currently its benefits are used
in systems like wireless computer networks, mobile communication networks, terrestrial digital television or optical fiber networks. The idea of the method is to encode
the information on a plurality of subcarriers, which are orthogonal on the time interval corresponding
to a single OFDM symbol. This results in immunity to destructive interference in multipath propagation,
greatly appreciated in indoor environments or cities, but involves specific problems, like
e.g. high peak-to-average power ratio and problems with nonlinear distortions. In particular, one of nonlinear
distortion sources can be digital-to-analog (D/A) as well as analog-to-digital (A/D) converters.

The process of D/A or A/D conversion of OFDM signal generally involves some inevitable signal degradation \cite{Colotti90, Maxim00, Kester04}.
It can be analyzed and characterized by means of numerical simulations, like in \cite{Come00, Lee05},
but although this way one can obtain precise results for the analyzed cases, obviously this approach does not provide knowledge of the system's behavior in general. For wider understanding
of the subject it is much more preferred to use an analytic model, showing explicitly relations
in the system. Some of such models have been already developed for signal clipping \cite{Gross94}, quantization
noise \cite{Mehrnia05} and their joint effects \cite{Dardari06, Berger11}, and one of important conclusions from them was that
degradation due to quantization noise can be reduced by oversampling the signal. However, it seems that little is still known about nonlinear
distortions in the conversion process, which may be important particularly for systems using high order modulations with constellations
densely populated by symbols. For this reason, the aim of this paper is to provide a rigorous, analytic model of pseudo-random
nonlinear distortions based on concepts described in \cite{Rudzinski10, Rudzinski12}. These two papers have laid
foundations for such theoretical model, introducing the idea of effective number of samples of a digital (discrete and quantized)
signal and in this paper this quantity is rigorously related to degradation of the signal.
The derived model shows, that like quantization noise, pseudo-random nonlinear distortions can be limited by
oversampling, but that this limiting is less efficient.

The organization of the paper is as follows. The definitions and assumptions used for the construction
of the model are presented in \secref{sec:def}. In \secref{sec:quantnoise} there is briefly discussed
the possibility of decreasing degradation by quantization noise by oversampling of the signal.
The assumed decomposition of nonlinear distortions into deterministic and pseudo-random parts
is introduced in \secref{sec:decomp}. Further, \secref{sec:tranprob} presents the derivation of transition
probability for OFDM signal, in \secref{sec:Neff} there is defined the effective number of samples
of the signal and an expression for this quantity is derived, and \secref{sec:pseudorandom} provides
the theoretical description of the signal's degradation by pseudo-random nonlinear distortions.
Finally, \secref{sec:summary} summarizes and concludes the paper.

\section{Definitions and assumptions}\label{sec:def}
The subject of the considerations here is a real, digital signal with OFDM coding, converted by a D/A or A/D converter
with the resolution of $n$ bits. The signal is formed by discretization and quantization of an ideal OFDM
symbol, given by superposition of $K$ modulated subcarriers:
\begin{equation}
x\od{t} = \sum_{k=1}^K A_k \cos\od{\omega_k t + \phi_k}.
\end{equation}
In general, individual subcarriers represent symbols from arbitrary constellations, defined by 
amplitudes $A_k$ and phases $\phi_k$, constant during the whole time interval $\TS$ of the OFDM symbol.
For simplicity, it is assumed, that all $A_k$ and $\phi_k$ are independent random variables, amplitudes $A_k \in \mathbb{R}_+$
and have the same probability distributions, while each $\phi_k$ satisfies $\mean{\e^{\I\phi_k}} = \mean{\e^{2\I\phi_k}} = 0$.
These assumptions are true for usually used quadrature modulations, like $M$-QAM and $M$-PSK with $M\geq 4$
(for other constellations, which do not meet the last assumptions, like BPSK, just some corrections of numerical
factors will be needed, still, the presented derivations remain essentially valid).
The OFDM symbol is assumed to have mean value equal to $0$ and angular frequencies are particularly chosen as
\begin{equation}
\omega_k = k \omega_1, \quad\text{with}\ \omega_1 = \frac{2\pi}{\TS},
\end{equation}
thus the modulated subcarriers are orthogonal on the time interval of the OFDM symbol. With these assumptions,
the mean square $\mean{A_k^2}$ is the same for each $k$, so that the mean power of the OFDM symbol is
\begin{equation}
\sigma^2 = K \frac{\mean{A_k^2}}{2}.\label{eq:sigma}
\end{equation}

The discretized form of the OFDM symbol is the sequence of $\NS$ samples
\begin{equation}
x_i\equiv x\od{iT}, \quad i = 0, 1, \ldots, \NS-1,
\end{equation}
with $i$ indexing consecutive samples and $T = \TS/\NS$ being the sampling period.
The discretization density (oversampling rate) is determined by the ratio $\NS/K$, assumed
to satisfy the Nyquist criterion: $\NS/K \geq 2$. It is known from the central limit theorem, that the signal $x_i$ can be well approximated
by Gaussian stochastic process with independent samples, taking value $x$ with probability density
dependent on the root mean square value \eqref{eq:sigma}:
\begin{equation}
\prob{x_i = x} = \frac{1}{\sqrt{2\pi\sigma^2}} \exp\od{-\frac{x^2}{2\sigma^2}}.\label{eq:probxi}
\end{equation}

Digital devices represent numbers with finite precision. In particular, the considered converter with resolution
of $n$ bits requires rounding (quantization) of $x_i$ values to integral numbers from the set $\Zn = \left\{-2^{n-1},\ldots,2^{n-1}-1\right\}$,
further called ``levels''. According to the above, the digital representation of the OFDM symbol is
the sequence of steps
\begin{equation}
\xD[i] = x_i + \DxQ_i, \quad \text{with}\ \xD[i]\in\mathbb{Z}_n: \abs{\xD[i]-x_i} = \min_{\xi\in\Zn} \abs{\xi - x_i},
\end{equation}
where $\DxQ_i$ is the quantization error (noise).

The highest and lowest values of the signal are limited, hence clipping of the signal
occurs \cite{Gross94, Dardari06, Berger11}. It is convenient to relate the power of the signal to the converter's dynamic range
(clipping level) with the help of the coefficient
\begin{equation}
\alpha = \frac{2^{n-1}}{\sigma}.
\end{equation}
The reasonable practical value is $\alpha \approx 4$. Further it is assumed here, that clipping is insignificant
compared to other degrading factors.

In real converters the levels differ from ideal values, what is characterized by so-called
differential and integral nonlinearity \cite{Maxim00}. This causes nonlinear distortions
of the converted signal. If the error for level $p$ is $\Dy\od{p}$,
then the signal with nonlinear distortions is
\begin{equation}
y_i = \xD[i] + \Dy\od{\xD[i]},
\end{equation}
where the sequence $\Dy\od{\xD[i]}$ depends on imperfections of the converter and also on the converted signal itself.
This has considerable consequences, qualitatively different than quantization. This results from the fact,
that unlike quantization noise, consecutive samples of the nonlinear distortions $\Dy\od{\xD[i]}$ can have
the same value with nonzero probability.

\section{Quantization noise}\label{sec:quantnoise}
It is known, that (using the introduced notation) if $\sigma \gg 1$, then the quantization noise $\DxQ_i$
is very well approximated by white noise with uniform distribution over the interval
$\left[-\frac{1}{2},\frac{1}{2}\right]$, i.e. of length the same as level separation \cite{Widrow08}.
In this case, essentially by definition, the whole power of the quantization noise
\begin{equation}
\mean{\bkt{\DxQ_i}^2} = \frac{1}{12}
\end{equation}
spreads equally over all the $\NS$ samples of spectrum, from which only $2K$ correspond to the signal.
Thus, the total power of the quantization noise in the signal band is
\begin{equation}
\sigmaq^2 = \frac{2K}{\NS} \mean{\bkt{\DxQ_i}^2} = \frac{1}{6} \frac{K}{\NS}.
\end{equation}
Hence, the signal-to-noise ratio for quantization noise:
\begin{equation}
\SNRq = \frac{\sigma^2}{\sigmaq^2} = \frac{3}{2} \bkt{\frac{2^n}{\alpha}}^2 \frac{\NS}{K},
\end{equation}
depends directly on oversampling factor $\NS/K$, scaling factor $\alpha$ and converter's resolution $n$.
In particular, this means, that the improvement of quantization noise can be obtained by increasing the number
of signal's samples and that the improvement of $\SNRq$ is simply proportional to that increase.

\section{Decomposition of nonlinear distortions}\label{sec:decomp}
The total error of converter level value can result from: constant offset and nonunitary gain, which shift
and change the slope of the converter's transient characteristics, and differential and integral nonlinearities \cite{Maxim00, Kester04}.
Offset and nonunitary gain cause no nonlinear distortions and can be relatively easily corrected. These two
error sources are further neglected here. Then, the remaining error results from nonlinearities and
can be decomposed into two components:
\begin{equation}
\Dy\od{p} = \Dy[d]\od{p} + \Dy[s]\od{p}.
\end{equation}
The first component $\Dy[d]\od{p}$ is slowly-varying or deterministic part of nonlinear distortions.
It can be regarded as systematic or regular deviation of converter's transient characteristic, approximately constant
within adjacent levels, i.e. related mainly to integral nonlinearity. The second component $\Dy[s]\od{p}$ is the quickly varying,
pseudo-random or stochastic term, representing irregular deviations without any clearly visible pattern. It is defined to have vanishing mean value:
\begin{equation}
\mean{\Dy[s]\od{p}} = 0
\end{equation}
and no correlation between errors of particular levels:
\begin{equation}
\mean{\Dy[s]\od{p}\Dy[s]\od{p'}} = \Dy[s]^2 \delta_{pp'}.\label{eq:meanDyspDyspp}
\end{equation}
The stochastic part is naturally associated to differential nonlinearity of A/D or D/A converter.
The model derived in this paper is taking this source of degradation into account.

Consecutive samples of pseudo-random nonlinear distortions corresponding to digital
signal $\xD[i]$ form a stochastic process that can be written down in the form:
\begin{equation}
\Dy[s]\od{\xD[i]} = \sum_{p\in\Zn} \Dy[s]\od{p} \delta_{p\xD[i]}.\label{eq:DysxDi}
\end{equation}
The autocorrelation function of this stochastic process is not trivial, despite of condition \eqref{eq:meanDyspDyspp}.
In fact, using \eqref{eq:DysxDi} and then \eqref{eq:meanDyspDyspp} one obtains:
\begin{equation}
R_i = \mean{\Dy[s]\od{\xD[0]} \Dy[s]\od{\xD[i]}} = \Dy[s]^2 \mean{ \delta_{\xD[0]\xD[i]} }.\label{eq:Ri}
\end{equation}
Obviously for $i=0$ the delta is equal to $1$ and hence $R_0 = \Dy[s]^2$.
For $i\neq 0$ and uncorrelated samples with values from a~continuous set, the above delta would determine
a zero-measure subset of the probability space, hence $R_i = 0$ for $i\neq 0$. This is the case for quantization
noise. However, samples of digital signal belong to discrete set $\Zn$, thus the probability of them being equal
is greater than zero and autocorrelation function of samples of pseudo-random nonlinear distortions
is more complicated. Its calculation is based on derivation of transition probability for the OFDM signal,
which is presented in the next section.

\section{Transition probability for digital OFDM signal}\label{sec:tranprob}
The transition probability of a signal is defined as the probability (or probability density for continuous case)
of observation of given values at two samples with given delay. Exploiting stationarity, the transition probability
for the digital signal $\xD[i]$ can be defined as:
\begin{equation}
\tprobq[i]{p}{p'} = \prob{\xD[0] = p \ \wedge\ \xD[i] = p'}.
\end{equation}
For analogue (with continuous values) case, i.e. signal $x_i$ before quantization, similarly is defined transition probability density:
\begin{equation}
\tprob[i]{p_0}{p_1} = \prob{x_0 = p_0 \ \wedge\ x_i = p_1}.
\end{equation}
These two functions are related by the obvious formula:
\begin{equation}
\tprobq[i]{p}{p'} = \lid{p_0}{p-\frac{1}{2}}{p+\frac{1}{2}} \lid{p_1}{p'-\frac{1}{2}}{p'+\frac{1}{2}} \tprob[i]{p_0}{p_1}.
\end{equation}
Probability of conjunction can be expressed by conditional probability, hence:
\begin{equation}
\tprob[i]{p_0}{p_1} = \probcon{x_i = p_1}{x_0 = p_0} \prob{x_0 = p_0}.
\end{equation}
The factor $\prob{x_0 = p_0}$ is given by \eqref{eq:probxi}, while the expression for
$\probcon{x_i = p_1}{x_0 = p_0}$ can be found starting from the definition:
\begin{equation}
w\od{t} = x\od{t} - x_0 = \sum_{k=1}^K \real{ A_k \e^{\I\phi_k} \bkt{ \e^{\I\omega_k t} - 1 } }.
\end{equation}
Then $\probcon{x_i = p_1}{x_0 = p_0} = \probcon{w\od{iT} = p_1 - p_0}{x_0 = p_0}$.
Because $w\od{iT}$ is a linear combination of multiple independent random variables, the central limit theorem
states, that its probability distribution can be well approximated by Gaussian:
\begin{equation}
\prob{w\od{iT} = x} = \frac{1}{\sqrt{2 \pi \sigmaw[i]^2}} \exp\od{-\frac{x^2}{2 \sigmaw[i]^2}},
\end{equation}
with variance
\begin{equation}
\sigmaw[i]^2 = \mean{w\od{iT}^2} = 2 \sigma^2 \bkt{ 1 - \frac{ \cos\frac{\bkt{K+1}\pi i}{\NS} \sin\frac{K \pi i}{\NS} }{ K \sin\frac{\pi i}{\NS} } } \label{eq:sigmawi2}
\end{equation}
(for $i=1$ it will be abbreviated $\sigmaw\equiv\sigmaw[1]$). It can be noted, that
$\lim_{i\rightarrow 0} \sigmaw[i] = 0$, just as it should follow from definition of $w\od{t}$.
Ignoring any statistical dependence between $w\od{iT}$ and $x_0$, the sought conditional probability
$\probcon{w\od{iT} = p_1 - p_0}{x_0 = p_0} \approx \prob{w\od{iT} = p_1 - p_0}$, then:
\begin{equation}
\tprob[i]{p_0}{p_1} \approx \frac{1}{2 \pi \sigmaw[i] \sigma} \exp\od{-\frac{\bkt{p_1 - p_0}^2}{2 \sigmaw[i]^2}} \exp\od{-\frac{p_0^2}{2\sigma^2}}
\end{equation}
and:
\begin{multline}
\tprobq[i]{p}{p'} = \frac{1}{2\pi \sigmaw[i] \sigma} \lid{p_0}{p-\frac{1}{2}}{p+\frac{1}{2}} \exp\od{-\frac{p_0^2}{2\sigma^2}}
\times \\
\lid{p_1}{p'-\frac{1}{2}}{p'+\frac{1}{2}} \exp\od{-\frac{\bkt{p_1 - p_0}^2}{2 \sigmaw[i]^2}}.
\end{multline}
This result can be rewritten using the error function
\begin{equation}
\erf{x} = \frac{2}{\sqrt{\pi}} \lid{t}{0}{x} e^{-t^2},
\end{equation}
and because $\sigma \gg 1$, it is a good approximation to assume that the interval of integration over $p_0$ is very narrow,
the integrand is essentially constant within these limits and evaluate it at $p_0 = p$. Then eventually:
\begin{multline}
\tprobq[i]{p}{p'} \approx \frac{1}{2 \sqrt{2 \pi \sigma^2}} \exp\od{-\frac{p^2}{2 \sigma^2}}
\times \\
\sbkt{ \erf{\frac{p'-p+\frac{1}{2}}{\sigmaw[i]\sqrt{2}}} - \erf{\frac{p'-p-\frac{1}{2}}{\sigmaw[i]\sqrt{2}}} }. \label{eq:tprob}
\end{multline}
Thus, the signal value is the same for both samples with probability
\begin{equation}
\tprobq[i]{p}{p} \approx \frac{1}{\sqrt{2\pi \sigma^2}} \exp\od{-\frac{p^2}{2 \sigma^2}} \erf{\frac{1}{2 \sigmaw[i]\sqrt{2}}}.
\end{equation}
It can be noted, that within the used approximations
\begin{equation}
\tprobq[i]{p}{p} \approx \frac{ \erf{\frac{1}{2 \sigmaw[i]\sqrt{2}}} }{ \erf{\frac{1}{2 \sigmaw\sqrt{2}}} } \tprobq[1]{p}{p} \label{eq:PippP1pp}
\end{equation}
and
\begin{equation}
\tprobq[0]{p}{p} \approx \frac{1}{\sqrt{2\pi \sigma^2}} \exp\od{-\frac{p^2}{2 \sigma^2}} = \prob{x_i = p}.
\end{equation}
Assuming, that summation over levels can be replaced by integration,
the total probability $\sum_p \tprobq[0]{p}{p} \approx 1$, meaning, that the derived expression is quite accurate.

Having found the transition probability, the next step is to calculate the effective number of samples,
defined further in the paper.

\section{Effective number of samples of digital OFDM signal}\label{sec:Neff}
\begin{figure*}[!tb]
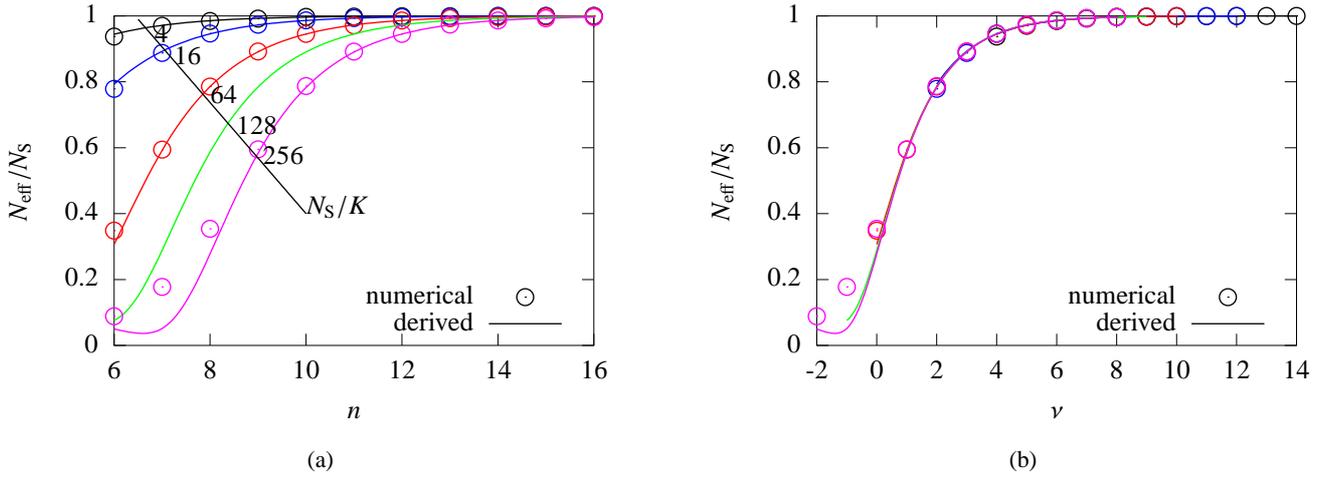

\begin{center}
\begin{subfigure}[c]{.5\textwidth}
\centering{
\input{calcNeff.tex}
\caption{}\label{rys:Neff_n}
}
\end{subfigure}~\begin{subfigure}[c]{.5\textwidth}
\centering{
\input{calcNeff_nzr.tex}
\caption{}\label{rys:Neff_nu}
}
\end{subfigure}
\caption{Normalized effective number of samples of OFDM signal $\Neff/\NS$ vs. (\subref{rys:Neff_n}) converter's resolution $n$ and (\subref{rys:Neff_nu}) reduced resolution $\nu$: comparison of numerical results from \cite{Rudzinski10} and the derived formula \eqref{eq:Neff_wynik}.}\label{rys:porownanie}
\end{center} 
\end{figure*}
Because values of $\xD[i]$ belong to a set with a finite number of elements, the probability that
two different samples have the same value is greater than zero. This causes, that in the sequence $\xD[i]$
appear constant subsequences of two or more samples, effectively reducing the number of samples of the signal.
Let $\mean{L}$ denote the mean length of constant subsequence within signal $\xD[i]$. The effective number
of samples of the signal $\xD[i]$ is hereby defined as:
\begin{equation}
\Neff = \frac{\NS}{\mean{L}}.
\end{equation}
Thus, $\Neff$ denotes the number of changes of values of consecutive samples in the digital signal.
Limiting values are: $\Neff = \NS$ if no sample is equal to the previous one, and $\Neff = 1$ if the signal
is constant (all samples are equal). It is convenient to use the normalized effective number of samples
$\Neff/\NS$, taking values from $1/\NS$ up to $1$.

From the definition it follows, that the effective number of samples can be calculated by counting
changes of values of consecutive samples. Therefore:
\begin{equation}
\Neff = \mean{ 1 + \sum_{i=1}^{\NS-1} \bkt{ 1 - \delta_{ \xD[i-1] \xD[i] } } }.
\end{equation}
For stationary stochastic process, mean value of the delta in this expression is the same for each pair
of samples, hence:
\begin{equation}
\frac{\Neff}{\NS} = 1 - \bkt{1-\frac{1}{\NS}} \mean{ \delta_{ \xD[0] \xD[1] } }. \label{eq:NeffNSmd01}
\end{equation}
The calculated transition probability allows to express the mean value of the Kronecker delta:
\begin{equation}
\mean{ \delta_{ \xD[0] \xD[1] } } = \sum_{p\in\Zn} \tprobq[1]{p}{p} \approx 2\sum_{p=0}^{2^{n-1}} \tprobq[1]{p}{p} - \tprobq[1]{0}{0}.
\end{equation}
Then:
\begin{multline}
\frac{\Neff}{\NS} \approx 1 - \frac{1}{\sqrt{2\pi \sigma^2}} \erf{\frac{1}{2 \sigmaw \sqrt{2}}}
\times \\
\sbkt{ 2 \sum_{p=0}^{2^{n-1}} \exp\od{-\frac{p^2}{2 \sigma^2}} -1 }.
\end{multline}
Again exploiting $\sigma\gg 1$ and approximating the summation over levels $p$ by integration,
one obtains the closed form expression:
\begin{equation}
\frac{\Neff}{\NS} \approx 1 - \erf{\frac{1}{2 \sigmaw\sqrt{2}}} \sbkt{ \erf{\frac{\alpha}{\sqrt{2}}} - \frac{1}{\sqrt{2\pi \sigma^2}} }.\label{eq:Neff_wynik}
\end{equation}
This formula is compared to results of numerical calculations shown in \figref{rys:porownanie}, plotted
against resolution $n$ and ``reduced resolution''
\begin{equation}
\nu = n - \log_2\frac{\NS}{K}.
\end{equation}
As it can be seen, the derived formula \eqref{eq:Neff_wynik} reproduces the effective number of samples very accurately in most cases.

The derived formula \eqref{eq:Neff_wynik} can be approximated to show more clearly relations between various parameters.
For the practically significant value of $\alpha\approx 4$ one has $\erf{\alpha/\sqrt{2}}\approx 1$.
The second term in brackets is inversely proportional to $\sigma$, thus quickly decreasing as $2^{-n}$. This term
is small and can be ignored. Expanding the trigonometric functions in $\sigmaw^2$ given by \eqref{eq:sigmawi2}
into Maclaurin series and then leaving only the leading term one obtains
\begin{equation}
\sigmaw \approx \frac{ 2^n \pi }{ \alpha \sqrt{3} } \bkt{\frac{\NS}{K}}^{-1} = \frac{ 2^\nu \pi }{ \alpha \sqrt{3} }.
\end{equation}
For small arguments $\erf{x} \approx 2 x /\sqrt{\pi}$, therefore for $\nu > 0$
the simplified formula is:
\begin{equation}
\frac{\Neff}{\NS} \approx 1 - \frac{ \alpha \sqrt{3} }{ 2^n \sqrt{2\pi^3}} \frac{\NS}{K} = 1 - \frac{ \alpha \sqrt{3} }{ 2^\nu \sqrt{2\pi^3} }.
\end{equation}
It can be seen that the only dependence on $n$ and $\NS/K$ is through the reduced resolution $\nu$, what explains
the behavior of results depicted in \figref{rys:Neff_nu}.

\section{Signal degradation by pseudo-random nonlinear distortions}\label{sec:pseudorandom}
\begin{figure}[tb]
\begin{center}
\input{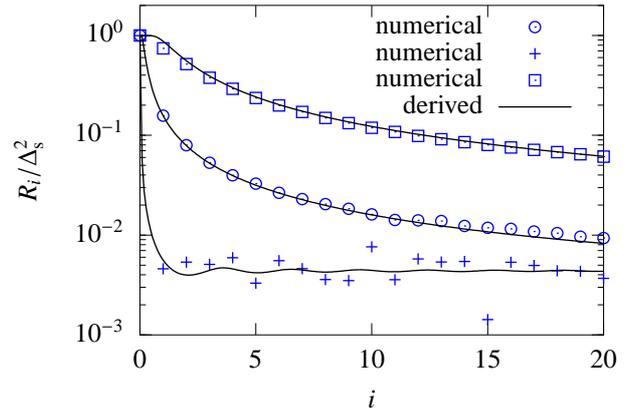}
\caption{Normalized autocorrelation function of pseudo-random nonlinear distortions: comparison of numerical results and the derived formula \eqref{eq:Rifinal} for three cases. Values are plotted only for a few smallest lags, which are the most interesting part of the autocorrelation function.}\label{rys:Rifinal}
\end{center} 
\end{figure}

The autocorrelation function $R_i$ given by \eqref{eq:Ri} contains mean value of Kronecker delta,
which can be calculated using the derived transition probability:
\begin{equation}
\mean{ \delta_{\xD[0]\xD[i]} } = \sum_{p\in\Zn} \tprobq[i]{p}{p}.
\end{equation}
Using \eqref{eq:PippP1pp} the mean value for samples $0$ and $i$ can be expressed by mean value
for samples $0$ and $1$:
\begin{equation}
\mean{ \delta_{\xD[0]\xD[i]} } = \frac{ \erf{\frac{1}{2 \sigmaw[i]\sqrt{2}}} }{ \erf{\frac{1}{2 \sigmaw\sqrt{2}}} } \mean{ \delta_{\xD[0]\xD[1]} }.
\end{equation}
Therefore, using \eqref{eq:NeffNSmd01} rearranged to express the mean value of delta by the effective number of samples,
one finds:
\begin{equation}
R_i \approx \bkt{ 1 - \frac{\Neff}{\NS} } \Dy[s]^2 \frac{ \erf{\frac{1}{2 \sigmaw[i]\sqrt{2}}} }{ \erf{\frac{1}{2 \sigmaw\sqrt{2}}} }.\label{eq:Rifinal}
\end{equation}
Exemplary results of averaged autocorrelation functions from numerical simulations and calculated with the help of the above formula are presented in \figref{rys:Rifinal}.
For each case the averaging has been conducted over 301 randomly generated signals and randomly generated level errors $\Dy\od{p}$.
It can be seen, that the derived expression correctly resembles the trends of the considered distortions.
It is interesting to note, that the autocorrelation function is generally a peak of certain width, which
can be characterized by the ratio
\begin{equation}
\frac{R_1}{R_0} = 1 - \frac{\Neff}{\NS},
\end{equation}
hence, the effective number of samples. It is the expected observation, since a lower effective number of samples
means more repetitions of values in the signal, thus higher correlation between consecutive samples.

\begin{figure}[tb]
\begin{center}
\input{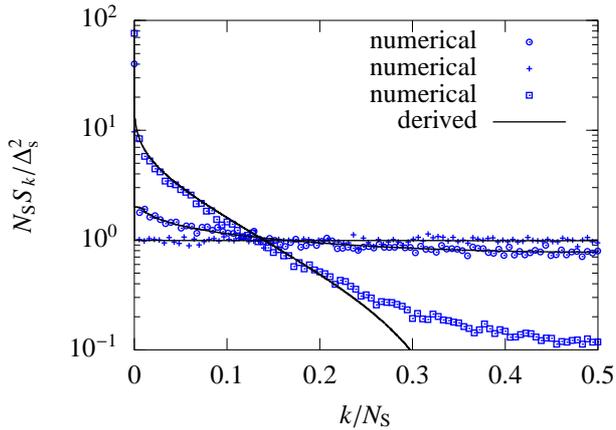}
\caption{Normalized power spectrum of pseudo-random nonlinear distortions: comparison of numerical results and the derived formula \eqref{eq:Sk} for three cases.}\label{rys:Sk}
\end{center} 
\end{figure}
From the Wiener-Khinchin theorem, the power spectrum of the pseudo-random distortions can be
calculated as Fourier transform of $R_i$:
\begin{multline}
S_k = \frac{1}{\NS}\sum_{i=0}^{\NS-1} R_i \e^{-\I\omega_k i T} =
\\
\frac{\Dy[s]^2}{\NS} \bkt{ 1 - \frac{\Neff}{\NS} } \sum_{i=0}^{\NS-1} \frac{ \erf{\frac{1}{2 \sigmaw[i]\sqrt{2}}} }{ \erf{\frac{1}{2 \sigmaw\sqrt{2}}} } \e^{-\I\omega_k i T}.\label{eq:Sk}
\end{multline}
The predictions given by \eqref{eq:Sk} are compared to averaged numerical results in \figref{rys:Sk}.
A good agreement is observed as the theoretical curves resemble the power spectra with high accuracy.
The only clearly visible discrepancy is for the frequencies with the lowest powers, however
it is possible, that the results would fit better if the number of results for averaging was increased.
Still, the total contribution of this part of spectrum is rather low and even if the derived expression is inaccurate
here, this will have no significant impact.

The total power of pseudo-random nonlinear distortions in the signal's band is
\begin{equation}
\sigmanls^2 = \sum_{\abs{k}\leq K, k\neq 0} S_k = \frac{1}{\NS} \sum_{i=0}^{\NS-1} R_i \frac{ \sin\frac{2\pi K i}{\NS} }{ \sin\frac{2\pi i}{\NS} }.\label{eq:sigmanls2}
\end{equation}
With $R_i$ given by the expression \eqref{eq:Rifinal} this leads to:
\begin{equation}
\sigmanls^2 = \frac{\Dy[s]^2}{\NS} \bkt{ 1 - \frac{\Neff}{\NS} } \sum_{i=0}^{\NS-1} \frac{ \erf{\frac{1}{2 \sigmaw[i]\sqrt{2}}} }{ \erf{\frac{1}{2 \sigmaw\sqrt{2}}} } \frac{ \sin\frac{2\pi K i}{\NS} }{ \sin\frac{2\pi i}{\NS} }.
\end{equation}
This expression for $\sigmanls^2$ is complicated and it would be beneficial to derive an approximation that, although less accurate,
would express the relations between quantities more clearly. Because $R_i$ is mainly a peak of certain width, it can be approximated in \eqref{eq:sigmanls2} with a triangular function:
\begin{equation}
R_i \approx
\begin{cases}
R_0 - \bkt{R_0 - R_1} i, & \abs{i} < \frac{\NS}{\Neff}, \\
0, & \text{otherwise},
\end{cases}
\end{equation}
where the range of $i$ is shifted by $-\NS/2$ -- this is possible because $R_i$ is periodic.
Then, nonzero terms of the sum are only near $i=0$, so one can use $\sin\bkt{2\pi i / \NS} \approx 2\pi i / \NS$
and the next step is replacement of summation with integration. However, in geometrical terms, since
in this case summations corresponds to adding areas of rectangles, while integration -- of triangles, and $R_0$
is expected to be the dominant component, to compensate for ignored area of the central step (peak) a factor
of $2$ is introduced: $\sum\rightarrow 2\int$. Hence, this way one obtains: 
\begin{multline}
\sigmanls^2 \approx \frac{2 \Dy[s]^2}{\pi} \sbkt{ \intsin\frac{2\pi K}{\Neff} - \frac{\Neff}{2\pi K} \bkt{ 1 - \cos\frac{2\pi K}{\Neff} } } =
\\
\frac{2 \Dy[s]^2}{\pi} \sbkt{ \frac{1}{1\cdot 2!} \frac{2\pi K}{\Neff} -\frac{1}{3\cdot 4!}\bkt{ \frac{2\pi K}{\Neff} }^3 +\ldots },
\end{multline}
where $\intsin x = x - x^3/(3\cdot 3!) + x^5/(5\cdot 5!) - x^7/(7\cdot 7!) + \ldots$ is the integral sine function.
Thus, in the rough approximation $\sigmanls^2 \approx 2 K \Dy[s]^2 / \Neff$ and the associated signal-to-noise ratio is
\begin{equation}
\SNRNL[s] = \frac{\sigma^2}{\sigmanls^2} \approx  \frac{1}{8 \Dy[s]^2} \bkt{\frac{2^n}{\alpha}}^2 \frac{\Neff}{K}.
\end{equation}
This result has the same form as for quantization noise with the effective number of samples $\Neff$ used in place
of total number of samples $\NS$.

\section{Summary}\label{sec:summary}
The expressions \eqref{eq:Rifinal} and \eqref{eq:Sk} presented in this paper accurately describe autocorrelation function and power spectrum
of distortions caused by pseudo-random, i.e. irregular, nonlinear distortions introduced by an A/D or D/A converter
into OFDM signal and this way extend the theoretical model of the conversion process present in the literature.
Their derivation is based on transition probability for OFDM signal and a newly introduced quantity,
the effective number of samples, for which the expressions \eqref{eq:tprob} and \eqref{eq:Neff_wynik} have been found.
It has been shown, that effective number of samples reveals certain information about the autocorrelation function
and power spectrum of the considered distortions. In general, this quantity is important for digital signals, because
it takes into account important stochastic property -- repetition of values of samples, which has significant consequences.
It is expected, that so defined effective number of samples should find more applications in description of digital signals.

\section*{Acknowledgments}
Dr. Krzysztof {\L}atuszy{\'n}ski has given some useful hints in the early stage of the presented derivations.
This work has been financially supported by the European Regional Development Fund within the framework of the
1.~priority axis of the Innovative Economy Operational Programme, 2007-2013, submeasure 1.1.2~``Strategic R\&D Research'',
contract \mbox{POIG.01.02.01-00-014/08}.

\bibliographystyle{spmpsci}
\bibliography{OFDM_ens_prnd_bib}

\end{document}